\documentstyle[12pt,twoside,fleqn,espcrc1,epsf]{article}

\newcommand{\AmS}{{\protect\the\textfont2
  A\kern-.1667em\lower.5ex\hbox{M}\kern-.125emS}}

\title{Reconstructing the Source in Heavy Ion Collisions from
       Particle Interferometry}

\author{Urs Achim Wiedemann, Boris Tom\'a\v{s}ik and Ulrich 
Heinz\address{Institut f\"ur Theoretische Physik,\\ 
              D-93040 Regensburg, Germany}%
          }

\begin{document}
\maketitle

\begin{abstract}
The preliminary CERN SPS NA49 Pb+Pb 158 GeV/A one- and two-particle
$h^-$-spectra at
mid-rapidity are consistent with a source of temperature $T\approx 130$ MeV, 
lifetime $\tau_0\approx 9$ fm/c, transverse flow $\eta_f \approx 0.35$, and a 
transverse geometric size which is twice as large as the cold
Pb nucleus.
\end{abstract}

\section{Introduction}
\label{sec1}

Hadronic one- and two-particle spectra provide for
each particle species information about the phase-space 
distribution $S(x,K)$ of the hadronic emission region~\cite{dronten}.
Once reconstructed from the measured spectra, $S(x,K)$
allows to distinguish between different dynamical scenarios of
heavy ion collisions. It provides an experiment-based
starting point for a dynamical back extrapolation into the hot
and dense early stage of the collision, where quarks and
gluons are expected to be the relevant physical degrees of 
freedom.

Our reconstruction of $S(x,K)$ is based on the hadronic one- and 
two-particle spectra
  \begin{eqnarray}
    E{dN\over d^3p} &=& \int d^4x\, S(x,p) =
     \textstyle{1\over 2\pi} {d^2N\over p_t\, dp_t\, dy}
        \lbrack 1 + 2\sum_{n=1}^\infty v_n\cos n(\phi -\psi_R) 
                \rbrack\, ,
    \label{1}\\
     C({\bf K},{\bf q})
      &=& 1 + {{\vert\int d^4x\, S(x,K)\, e^{i q\cdot x}\vert^2}
              \over
              { \int d^4x S(x,p_1)\, \int d^4y\, S(y,p_2)}}
            = 1 + \lambda\exp\left[ 
              -\sum_{ij} R_{ij}^2({\bf K})\, {\bf q}_i\, {\bf q}_j
              \right]\, .
   \label{2}
  \end{eqnarray}      
The triple-differential hadronic one-particle spectrum (\ref{1})
tests the momentum-dependence of $S(x,K)$ only.
Its azimuthal $\phi$-dependence with respect to the reaction plane
$\psi_R$ is parametrized by the harmonic coefficients $v_n$~\cite{volo}. 
Information about the space-time structure of $S(x,K)$ can be obtained 
from the relative momentum dependence of the two-particle
correlator $C({\bf K},{\bf q})$,
${\bf q} = {\bf p}_1-{\bf p}_2$. This ${\bf q}$-dependence is usually
parametrized via the Hanbury-Brown Twiss (HBT) radius parameters
$R_{ij}^2({\bf K})$ which depend on the average pair momentum
${\bf K} = \textstyle{1\over 2}({\bf p}_1+{\bf p}_2)$. Depending
on the Gaussian parametrization adopted, the indices
$i$, $j$ in (\ref{2}) run either over the Cartesian directions
{\it long} (parallel to the beam), {\it out} (parallel to the
transverse component ${\bf K}_{\perp}$) and {\it side}, or over
the corresponding Yano-Koonin coordinates $q_{\perp} =
\sqrt{q_o^2 + q_s^2}$, $q_3$ and $q^0 = {\bf q}\cdot {\bf K}/K_0$
~\cite{dronten}.
%
\begin{figure}[h]\epsfxsize=13cm 
\centerline{\epsfbox{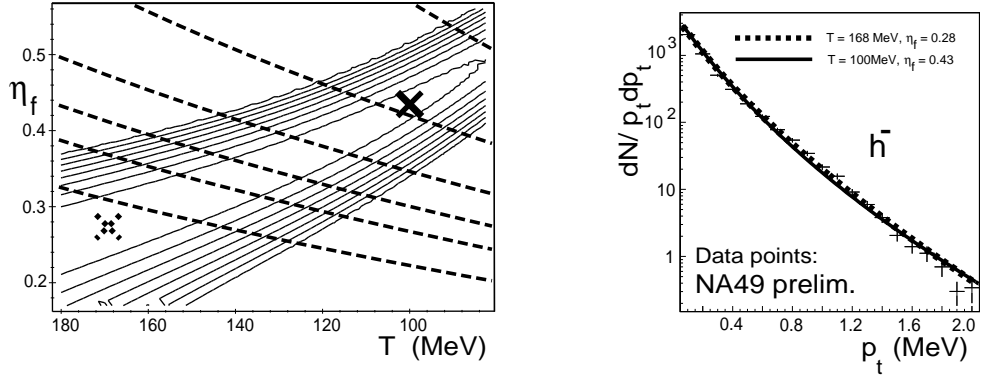}}
\vspace{-1cm}
\caption{LHS: $\chi^2$ contour plot of a fit to the NA49~\cite{jones}
$h^-$-spectrum. Dashed lines are for constant values of $\eta_f^2/T$.
RHS: different combinations of temperature $T$ and transverse flow
$\eta_f$ account for the same one-particle slope.
}
\vspace{-0.5cm}
\label{fig1}
\end{figure}
%
\section{Reconstructing an azimuthally symmetric source}
\label{sec2}

A typical data analysis starts from a simple ansatz
for the phase space distribution $S(x,K)$ in terms of very
few, physically intuitive fit parameters,~\cite{dronten,reson}:
  \begin{eqnarray}
     S_{\pi}(x,p) &=& S_{\pi}^{\rm dir}(x,p) + \sum_{R} S_{R\to \pi}(x,p)\, ,
  \label{3} \\
   S^{\rm dir}_i(x,P) &=& \textstyle{2J_i + 1 \over (2\pi)^3}\,
   P{\cdot}n(x)\,
   \exp{\left(- {P \cdot u(x) \over T} \right)}\,  
          \exp\left( - {r^2\over 2 R^2} 
                     - {\eta^2\over 2 (\Delta\eta)^2}
                     - {(\tau-\tau_0)^2 \over 2 (\Delta\tau)^2}
                 \right) \, . 
                 \label{4} 
  \end{eqnarray}
This model e.g. assumes local thermalization at freeze-out with 
temperature $T$ within a space-time region of transverse Gaussian 
width $R$, emission duration $\Delta\tau$, longitudinal extension 
$\tau_0\Delta\eta$, where $\eta= {1\over 2} \ln{[(t+z)/(t-z)]}$,
and average emission time $\tau_0$. 
The model allows for dynamical source correlations via the 
hydrodynamic flow $4$-velocity $u_{\mu}(x)$. We assume a linear
transverse flow profile $\eta_t(r) = \eta_f \left({r\over R}\right)$
with variable strength $\eta_f$, and Bjorken scaling of the flow
component in the longitudinal direction, $v_l=z/t$,
$\eta_l \equiv {1 \over 2} \ln[(1 + v_l)/(1-v_l)] = \eta$.
Resonances are produced in thermal abundances with proper 
spin degeneracy $2J_i+1$ for each particle species $i$. Their
contribution to the pion yield is obtained by propagating
them along their classical path $x^\mu = X^\mu + {P^\mu\over M} \tau$ 
according to an exponential decay law~\cite{reson},
  \begin{equation}
   S_{R\to\pi}(x,p) = 
        \int_{\bf R}\, \int d^4X \, 
        \int d\tau \, \Gamma e^{-\Gamma\tau} \,
        \delta^{(4)}\textstyle{\left[ x - 
            \left( X + {P\over M} \tau \right) \right]} 
        S_R^{\rm dir}(X,P)\, ,
 \label{5}
 \end{equation}
where $\int_{\bf R}$ is the integral over the available resonance
phase space for isotropic decays.
We include all pion decay channels of $\rho$, $\Delta$, $K^*$, 
$\Sigma^*$, $\omega$, $\eta$, $\eta'$, $K_S^0$, $\Sigma$ and $\Lambda$ 
with branching ratios larger than 5 percent.
\begin{figure}[h]\epsfxsize=16cm 
\centerline{\epsfbox{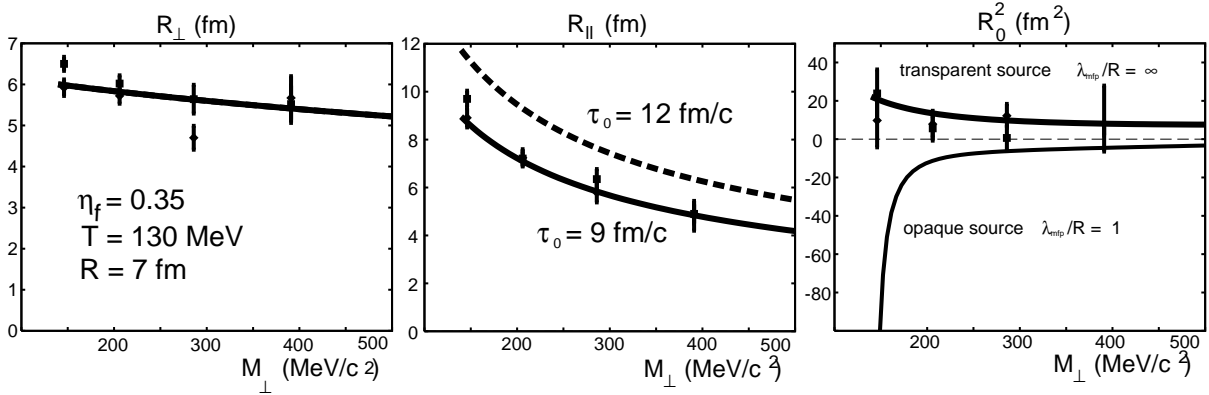}}
\vspace{-1cm}
\caption{Yano-Koonin-Podgoretskii HBT-radius parameters. The preliminary
NA49 Pb+Pb data~\cite{appels} are for $h^+h^+$ (squares) and $h^-h^-$ 
(diamonds) correlations.
}
\label{fig2}
\end{figure}
The model parameters $T$, $\eta_f$, $R$, $\Delta\eta$, 
$\Delta\tau$, $\tau_0$ can then
be determined via the following strategy:
  \begin{enumerate}
    \item
      {\it transverse one-particle spectrum $dN/dM_{\perp}^2$ determines
        blue-shifted temperature $T_{\rm eff}$:}\\
      The slope of $dN/dM_{\perp}^2$ is essentially given by
      $T_{\rm eff} = T \sqrt{ (1 + \eta_f)/ (1 - \eta_f)}$. 
      Hence, different combinations of $T$ and $\eta_f$ can 
      account for the same data, see Fig.~\ref{fig1}. 
    \item
      {\it $dN/dM_{\perp}^2$ \underline{and} $R_{\perp}(M_{\perp})$ 
        disentangle temperature and transverse flow. $R_{\perp}$ 
        fixes transverse extension $R$:}\\
      The $M_{\perp}$-slope of $R_{\perp}$ is proportional to $\eta_f^2/T$,
      $R_{\perp}^2 \approx R^2/(1 + M_{\perp}\eta_f^2/T)$ \cite{chap,WSH}.
      This slope allows in combination with $dN/dM_{\perp}^2$ to 
      specify $T$ and $\eta_f$, see Fig.~\ref{fig2}.
      The overall size of $R_{\perp}$ determines the Gaussian width $R$.
      We find $\eta_f \approx 0.35$, $T \approx 130$ MeV, $R \approx 7$ fm.
    \item
      {\it $R_{\parallel}$ determines $\tau_0$, width of
        rapidity distribution $dN/dY$ determines $\Delta\eta$:} \\
      In principle, $R_{\parallel}$ depends on $\tau_0$, $\Delta\eta$
      and $\Delta\tau$ \cite{WSH}. We have fixed $\Delta\eta = 1.3$ 
      by matching the width of the pion
      rapidity distribution. The data then favour clearly a
      lifetime of $\tau_0 \approx 9$ fm/c (see Fig.~\ref{fig2}) but are 
      not very sensitive to the emission duration $\Delta\tau$~\cite{boris}.
      The present plots are obtained with
      $\Delta\tau = 1.5$ fm/c.
    \item
      {\it $R_0$ discards opaque sources}:\\
      For the model (\ref{3}-\ref{5}), the YKP-parameter $R_0$ is
      mainly sensitive to the temporal aspects of the source. 
      The large statistical
      uncertainties for $R_0$ do not allow to constrain the model 
      parameter space further, see Fig.~\ref{fig2}. Models of opaque 
      sources including an opacity factor in (\ref{4})
      are excluded already by the present data~\cite{boris}.
  \end{enumerate}    
The radius of a cold Pb nucleus corresponds to $R_{\rm Pb}({\rm cold}) 
\approx 3.5$ fm
in the Gaussian parametrization (\ref{4}), i.e., the experimental
data indicate a very large source 
$R \approx 2\, R_{\rm Pb}({\rm cold})$. This is dynamically
consistent with a collision scenario in which the initially produced
pressure gradients result in a significant transverse flow 
$\eta_f = 0.35$, driving the expansion of the system
over a time of 9 fm/c to twice its initial size. The temperature may have
decreased substantially during this expansion; the data indicate 130 MeV at
freeze-out. These conclusions are further supported by the analysis
presented by G. Roland~\cite{roland}, which is based on approximate
analytical formulas. We have extracted
the model parameters by comparison to numerical model calculations,
following the above strategy. They are not obtained
from a simultaneous fit to all observables, and hence
we do not quote errors.

\section{Particle interferometry for collisions with finite
impact parameter}
\label{sec3}

The reaction plane analysis of triple-differential one-particle
spectra (\ref{1}) has been discussed extensively in this
conference e.g. by A. Poskanzer, J.-Y. Ollitrault, and S. Voloshin.
The general strategy for linking this analysis to an azimuthally
sensitive particle interferometry is based on the $\Phi$-dependence 
of the HBT-radii~\cite{aniso}, where $\Phi$ measures the azimuthal angle
of $\vec{\bf K}_{\perp}$ relative to the reaction plane,
  \begin{equation}
    R_{ij}^2(K_\perp,\Phi,Y) = 
             R_{ij,0}^2(K_\perp,Y)
             + 2 \sum_{n=1}^\infty
               {R^{c}_{ij,n}}^2(K_\perp,Y) \cos n\Phi
             + 2 \sum_{n=1}^\infty
               {R^{s}_{ij,n}}^2(K_\perp,Y) \sin n\Phi\, .
     \label{6}
  \end{equation}
Here, the many harmonic coefficients $R^{*}_{ij,*}$ 
make a direct comparison to experimental data impossible. However,
various relations hold amongst these coefficients, since the
leading anisotropy in realistic source models $S(x,K)$ can be
quantified by very few parameters only. Using the symmetries of the
system and assuming that elliptic deformations dominate we 
find~\cite{aniso}
  \begin{eqnarray}
    0 &\approx& {R_{l,m}^c}^2 \approx {R_{l,m}^s}^2 \approx
    {R_{ol,m}^c}^2 \approx {R_{ol,m}^s}^2 \approx 
    {R_{sl,m}^c}^2 \approx {R_{sl,m}^s}^2 \qquad \, ,m\geq 1\, ,
    \label{7} \\
    \alpha_1 &\approx&
             {1\over 3}{R_{o,1}^c}^2\, \approx\,
             {R_{s,1}^c}^2\, \approx\, 
             - {R_{os,1}^s}^2\, ,
             \label{8}\\
    \alpha_2 &\approx&
            -{R_{o,2}^c}^2\, \approx {R_{s,2}^c}^2\, \approx\, 
            {R_{os,2}^s}^2 \, .
            \label{9}
  \end{eqnarray}
A violation of Eqs. (\ref{8})-(\ref{9}) by experiment would indicate
strong higher order deformations and rule out many model scenarios.
On the basis of Eqs. (\ref{7})-(\ref{9}),
an azimuthally sensitive parametrization of the two-particle
correlator involves only two additional fit parameters,
  \begin{eqnarray}
    C_{\psi_R}({\bf K},{\bf q}) &\approx& 1 + 
        \lambda \exp\lbrack
        -{R_{o,0}}^2\, q_o^2 - {R_{s,0}}^2\, q_s^2
             - {R_{l,0}}^2\, q_l^2 - 2\, {R_{ol,0}}^2\, q_o q_l \rbrack\,
        \nonumber \\
        && \qquad 
          \times \exp\big\lbrack -\alpha_1\,
                (3\, q_o^2 + q_s^2)\, \cos(\Phi-\psi_R)\,
               + 2\alpha_1\, q_o q_s \sin(\Phi -\psi_R) \big\rbrack
        \nonumber \\
        && \qquad 
          \times \exp\big\lbrack -\alpha_2
                (q_o^2-q_s^2)\cos 2(\Phi-\psi_R)
                +2\,\alpha_2 q_oq_s \sin 2(\Phi-\psi_R) \big\rbrack\, .
    \label{10}
  \end{eqnarray}
The anisotropy parameter $\alpha_1$ vanishes at mid-rapidity or
if the source contains no dynamical correlations. It characterizes
anisotropic dynamics. The parameter $\alpha_2$ characterizes
the elliptic geometry. The parameters $\alpha_1$ and $\alpha_2$ 
can be determined from event samples in spite of the uncertainty in the
eventwise reconstruction of the reaction plane. For details,
see Ref.~\cite{aniso}.

This work is supported by BMBF, DAAD, DFG and GSI.

\end{document}